\documentclass[12pt]{article}
\usepackage{paper2e}
\usepackage{mydefs2e}
\usepackage{epsf}

\renewcommand{\Yasymm}{\raisebox{-3pt}{\drawsquare{7}{0.6}\hskip-7.6pt%
\raisebox{7pt}{\drawsquare{7}{0.6}}}}

\newcommand{\Yone}{\hbox{\bf 1}}

\begin{document}
\begin{titlepage}
\preprint{hep-ph/9709306\\
UMD-PP-98-98\\
UCB-PTH-97/43}

\title{New Mechanisms of\\\medskip
Dynamical Supersymmetry Breaking\\\medskip
and Direct Gauge Mediation}

\author{Markus A.~Luty\footnote{Sloan Fellow}}

\address{Department of Physics,
University of Maryland\\
College Park, Maryland 20742, USA\\
{\tt mluty@physics.umd.edu}}

\author{John Terning}

\address{Department of Physics,
University of California\\
Berkeley, California 94720, USA\\
{\rm and} \\
Theory Group,
Lawrence Berkeley National Laboratory\\
Berkeley, CA 94720, USA\\
{\tt terning@alvin.lbl.gov}}

\begin{abstract}
We construct \susc gauge theories with new mechanisms of
dynamical \susy breaking.
The models have flat directions at the classical level,
and different mechanisms lift these flat directions in different
regions of the classical moduli space.
In one branch of the moduli space,
\susy is broken by confinement in a novel manner.
The models contain only dimensionless
couplings and have large groups of unbroken global symmetries,
making them potentially interesting for model-building.
As an illustrative application, we couple the standard model gauge
group to a model with an $\SU{5}$ global symmetry,
resulting in a model with composite messengers and a
non-minimal spectrum of superpartner masses.
\end{abstract}

\date{September, 1997}

\end{titlepage}

\section{Introduction}
The last few years have seen a revival of interest in models in which
\susy is broken at low energy scales \cite{OldGaugeMed,NewGaugeMed}.
In this work, there has been a fruitful interplay between
theoretical progress in understanding dynamical \susy breaking
\cite{OldDynSUSYBreak,threetwo,NewDynSUSYBreak,ISS,IT} and model-building
(for recent progress in gauge-mediated
model-building, see
\eg \Refs{PTmodel,hitoshidirect,CERNdirect,mldirect,randall}).

In this paper, we construct a class of models that exhibit a new
mechanism of \susy breaking.
In these models, there is a classical flat direction that can
be parameterized by a composite ``baryon'' chiral superfield
$B \sim Q^N$, where $Q$ is an elementary chiral superfield.
This field gets a dynamical superpotential
\beq
W_{\rm dyn} \sim B^{p} \sim Q^{N p}.
\eeq
For large $Q$, the \Kahler potential is approximately canonical in
$Q$, so if $N p > 1$ the potential for $B$ slopes toward $B = 0$.
For small $B$, the models exhibit smooth confinement
(``s-confinement'') \cite{seiberg,sconfine},
and the \Kahler potential is smooth in $B$.
In this case, if $p < 1$ the potential for $B$ slopes away from
$B = 0$.
Since the vacuum energy does not vanish for any value of $B$, \susy
is broken with $\avg{B} \ne 0$.%
\footnote{The model of \Ref{ISS} also breaks \susy by confinement,
but that model has a linear potential at the origin.}

The models considered here have additional classical flat directions
as well as large groups of global symmetries.
We are able to obtain a great deal of information about the location
of the global minimum in the field space, but some important
properties of the ground state depend on non-calculable strong dynamics.

We then use the models constructed above as building blocks for
realistic models of gauge-mediated \susy breaking.
We construct an illustrative example by gauging a global
symmetry with the standard-model gauge group.
The resulting model has composite fermions that are charged under the
standard-model gauge group, and we add additional interactions so
that the composite fermions obtain a Dirac mass
with elementary fields.
This model can be realistic, and gives rise to interesting
phenomenology.
(For the model to work, we must make some assumptions
about the signs of non-calculable \Kahler terms, and
the \susy-breaking masses are also non-calculable.)

This paper is organized as follows.
In Section 2, we describe models that realize the \susy-breaking
mechanism described above.
In Section 3, we construct gauge-mediated \susy breaking models.
Section 4 contains our conclusions.
Some additional \susy-breaking models related to the models
discussed in Section 2 are analyzed in the Appendix.
These models also have classical flat directions and break
\susy through novel mechanisms.

\section{$\Sp{2N} \times \SU{2N - 1}$ models}
In this Section, we analyze models with gauge and global symmetry
group%
\footnote{In our conventions, the fundamental \rep of $\Sp{2N}$
has dimension $2N$.}
\beq
G = \Sp{2N} \times \SU{2N - 1} \times
[ \SU{2N - 1} \times \U1 \times \U1_R ],
\eeq
where the global symmetries are written in brackets.
The matter content is
\beq\bal
Q &\sim (\Yfund, \Yfund) \times (\Yone; 1, 1),
\\
L &\sim (\Yfund, \Yone)
\times (\Yfund; -1, -\sfrac{3}{2N - 1}),
\\
\bar{U} &\sim (\Yone, \bar{\Yfund}) \times
(\bar{\Yfund}; 0, \sfrac{2N + 2}{2N - 1}),
\\
\bar{D} &\sim (\Yone, \bar{\Yfund}) \times (\Yone; -6, -4N),
\eal\eeq
and there is a tree-level superpotential
\beq
W = \la Q L \bar{U}.
\eeq
The field content and superpotential of this model are reminiscent
of the ``3--2'' model of dynamical \susy breaking \cite{threetwo}.
(In fact we will see that the dynamics is similar to that of the
3--2 model in one branch of the moduli space.)
If we turn off the $\Sp{2N}$ gauge coupling and the superpotential,
$\SU{2N - 1}$ s-confines for any $N \ge 2$ \cite{seiberg}.
If we turn off the $\SU{2N - 1}$ gauge coupling and the superpotential,
$\Sp{2N}$ is in a non-Abelian Coulomb phase for $N \ge 6$,
it has a weakly-coupled dual description for $N = 4,5$,
it s-confines for $N = 3$, and confines with a quantum-deformed moduli
space for $N = 2$ \cite{seiberg,seibergdual}.

If we include the effects of the tree-level superpotential,
this theory has a classical moduli space that can be parameterized by
the gauge-invariants
\beq\bal
M_{LL} &= L L
\sim (\Yasymm; -2, -\sfrac{6}{2N - 1}),
\\
\bar{B}_U &= \bar{U}^{2N - 2} \bar{D}
\sim (\Yfund; -6, -\sfrac{4(N^2 - N + 1)}{2N - 1}),
\\
\bar{B}_D &= \bar{U}^{2N - 1}
\sim (\Yone; 0, 2N + 2),
\eal\eeq
subject to the constraints
\beq
(M_{LL})^{jk} (\bar{B}_U)^\ell \ep_{k\ell m_1 \cdots m_{2N - 3}} = 0,
\qquad
(M_{LL})^{jk} \bar{B}_D = 0.
\eeq
These constraints split the moduli space into two branches:
on one of them $M_{LL} = 0$ and $\bar{B}_U, \bar{B}_D \ne 0$,
and on the other $M_{LL} \ne 0$ and $\bar{B}_U, \bar{B}_D = 0$.

\subsection{The ``Baryon'' Branch}
We first consider the branch where $\bar{B}_U, \bar{B}_D \ne 0$.
In terms of the elementary fields, this corresponds to
the \vevs (up to gauge and flavor transformations)
\beq
\avg{\bar{U}} = \pmatrix{ v \cos\theta & \cr
& v \Yone_{2N - 2} \cr},
\qquad
\avg{\bar{D}} = \pmatrix{ v \sin\theta \cr 0 \cr \vdots \cr 0 \cr},
\eeq
where $\Yone_{2N - 2}$ is the $(2N - 2)$-dimensional identity matrix.
Far out along this flat direction, the $\SU{2N - 1}$ gauge group is
completely broken, and the fields $Q$ and $L$ get masses of order
$\la v$ (for $\cos\th \ne 0$).
Below the scale $\la v$, the effective theory is $\Sp{2N}$ super
Yang--Mills, and gaugino condensation in this theory gives rise to
the dynamical superpotential
\beq[Wdyn]
W_{\rm eff} \simeq \frac{\La_{\rm Sp}^3}{16\pi^2} \left(
\frac{4\pi \la \bar{U}}{\La_{\rm Sp}}
 \right)^{(2N - 1)/(N + 1)}.
\eeq
For $v \gg \La_{\rm SU}$, the \Kahler potential is
approximately canonical in $\bar{U}$, and so the potential for
$\bar{U}$ slopes toward $\bar{U} = 0$ for $N > 2$.
(The special case $N = 2$ will be considered separately below.)
However, if $\bar{U}$ becomes small, we must reconsider the
analysis.%
\footnote{The analysis for the case $\cos\th = 0$ is somewhat different.
In that case, the $\Sp{2N}$ theory has one light
flavor that would run away if there were no other interactions.
However, the runaway direction is not $D$ flat, and so there is
no \susc vacuum with $\cos\th = 0$.}

The physics for small field values depends on the relative strength
of the two gauge groups.
We first consider $\La_{\rm SU} \gg \La_{\rm Sp}$.
(This is the situation that would arise if the two groups were unified
at a higher scale.)
In this case, the analysis above breaks down for
$v \lsim \La_{\rm SU} / (4 \pi)$, the scale at which the massive
$\SU{2N - 1}$ gauge bosons have mass $g_{\rm SU}\, v \sim \La_{\rm SU}$
according to ``\naive dimensional analysis'' \cite{fourpi}.
For small values of $\avg{\bar{B}_D}$, we can use a description where
$\SU{2N - 1}$ s-confines, and we obtain an effective theory
(after integrating out states with mass $\sim \la \La_{\rm SU} / (4\pi)$)
with symmetry group
\beq
G_{\rm eff} = \Sp{2N} \times \left[ \SU{2N - 1}
\times \U1 \times \U1_R \right],
\eeq
matter content
\beq\bal
M_{QD} &= Q \bar{D} \sim \Yfund \times (\Yone; -5, -4N + 1),
\\
B_Q &= Q^{2N - 1} \sim \Yfund \times (\Yone; 2N - 1, 2N - 1),
\\
\bar{B}_U &= \bar{U}^{2N - 2} \bar{D}
\sim (\Yfund; -6, -\sfrac{4(N^2 - N + 1)}{2N - 1}),
\\
\bar{B}_D &= \bar{U}^{2N - 1}
\sim (\Yone; 0, 2N + 2),
\eal\eeq
and an effective superpotential
\beq
W_{\rm eff} = B_Q M_{QD} \bar{B}_D.
\eeq
If this were an elementary theory, the $\Sp{2N}$ dynamics would force
$\bar{B}_D$ to run away.
This can again be described by a superpotential of the form of \Eq{Wdyn},
but in the regime we are now considering the \Kahler potential is smooth
in the field $\bar{B}_D$.
Because the field $\bar{B}_D$ is composite, we know that if
$\avg{\bar{B}_D}$ is large compared to $\La_{\rm SU}$, we should
use the previous analysis in terms of the elementary degrees of freedom.
But this analysis shows that there is no \susc vacuum for large
field values, and we conclude that \susy is broken.
We see that this model realizes the mechanism of \susy breaking
described in the Introduction.

Note that the considerations above imply that there must be at least
a local \susy-breaking minimum with $\avg{\bar{B}_D} \ne 0$, since
there are no classical flat directions that can connect this vacuum
to the other branch of the moduli space.
The \susy-breaking order parameter is
\beq[vacf]
F \simeq \frac{\la^{(2N - 1) / (N + 1)} \La_{\rm Sp} \La_{\rm SU}}{4 \pi}
\left( \frac{\La_{\rm Sp}}{\La_{\rm SU}} \right)^{3 / (N + 1)}.
\eeq

We see that this model has two descriptions:
a ``Higgs'' description in which the gauge group $\SU{2N - 1}$ is broken,
and a ``confining'' description in which it confines.
This model therefore realizes the ``complementarity'' picture
described in \Refs{compliment}.
Neither of these descriptions is quantitatively under control
near the vacuum of the theory, but both pictures should be a reliable
guide to qualitative features of the low-energy physics.
We are not able to determine whether or not $\avg{\bar{B}_U}$
is nonzero.
(This can be thought of as the question of whether the induced soft
mass-squared for $\bar{B}_U$ is positive or negative at $\bar{B}_U = 0$.)
If $\avg{\bar{B}_U} = 0$, the global symmetry is broken down to
$\SU{2N - 1} \times \U1$, and there is a massless composite fermion
\beq
\psi \sim (\bar{\Yfund}; -6).
\eeq
If $\avg{\bar{B}_U} \ne 0$, the global symmetry is broken down to
$\SU{2N - 2} \times \U1$ (where the unbroken $\U1$ is a linear combination
of the original $\U1$ and a broken $\SU{2N - 1}$ generator), and there
are massless composite fermions
\beq
\psi \sim (\Yfund; -30),
\qquad
\chi \sim (\Yone; 0).
\eeq
In the confined description, the composite fermions correspond to the
fermion components of $\bar{B}_D$, and in the Higgs description they
correspond to the fermion component of $\bar{D}$.

It is amusing that the model above does not have gauge anomalies if we
replace $\Sp{2N}$ by either $\SU{2N}$ or $\SO{2N}$.
The $SO$ model breaks \susy by a mechanism very similar to the one
described above, but the $SU$ model does not break \susy!
The reason is that the
analog of the dynamical superpotential \Eq{Wdyn} in the $SU$ model is
\beq
W_{\rm eff} \sim \bar{U}^{(2N - 1) / (2N)},
\eeq
which gives rise to a potential that runs away for large $\bar{U}$.
We will not analyze the $SO$ version of the model in this paper.

We now briefly consider the analysis for small field values when
$\La_{\rm Sp} \gg \La_{\rm SU}$.
The analysis depends on the value of $N$.

For $N = 2$, the $\Sp{4}$ group has a confined description with a
deformed moduli space.
The tree-level superpotential turns into a mass term that combines
with the quantum constraint to force some of the composite fields in
this description to run away.
This shows that there is no \susc vacuum for small fields in
this model.

For $N = 3$, the $\Sp{6}$ group s-confines, and the low-energy theory is
an $\SU{5}$ gauge theory with matter content $\bar{\Yfund} \oplus \Yasymm$
plus singlets.
This theory is known to break \susy \cite{fivebarten},
so there is no \susc vacuum in this model for small fields.
This mechanism leads to a class of models that are discussed in the
Appendix.

For $N \ge 4$, the $\Sp{2N}$ group has a dual description in terms of a
$\Sp{2N - 6}$ gauge group.
The $\SU{2N - 1}$ matter content is
$\Yasymm \oplus (2N - 5) \cdot \bar{\Yfund}$ plus singlets.
This theory has a dynamically-generated superpotential \cite{PTasymm},
and this combines with the tree-level superpotential to give a runaway
behavior.
This again shows that there is no \susc vacuum for small fields.

\subsection{The ``Lepton'' Branch}
We now consider the branch where $\avg{L} \ne 0$.
Along this branch, we have
\beq
\avg{L} = \pmatrix{ v_1 \Yone_2 & & & \cr
& \ddots & & \cr
& & v_{N - 1} \Yone_2 & \cr
& & & 0 \cr & & & 0 \cr}.
\eeq
Ignoring global $\U1$ factors, this breaks the gauge and flavor symmetries
down to
\beq
G_{\rm eff}
= \SU{2} \times \SU{2N - 1} \times \left[ \SU{2}^{N - 1} \right],
\eeq
with light fields
\beq[light]
\bal
L' &\sim (\Yfund, \Yone) \times \Yone,
\\
Q' &\sim (\Yfund, \Yfund) \times \Yone,
\\
\bar{U}' &\sim (\Yone, \bar{\Yfund}) \times \Yone,
\\
\bar{D}' &\sim (\Yone, \bar{\Yfund}) \times \Yone,
\\
L^{\prime\prime} &\sim (\Yone, \Yone) \times \Yasymm,
\\
L^{\prime\prime\prime} &\sim (\Yone, \Yone) \times \Yfund,
\eal\eeq
and superpotential
\beq
W = \la Q' L' \bar{U}'.
\eeq
(Each $\SU{2}^{N - 1}$ \rep is denoted by a $\SU{2N - 2}$ \rep that
is understood to be decomposed under $\SU{2N - 2} \to \SU{2}^{N - 1}$.)
The only flat directions are excitations of $L$, which correspond to
the fields $L^{\prime\prime}$ and $L^{\prime\prime\prime}$ in \Eq{light}.
The remaining light fields have quartic potentials from the $D$-term
potential.

We will assume that the $\SU{2N - 1}$ group in the effective theory above
is stronger than the $\SU{2}$.
This is always true for $N \ge 6$, where the $\SU{2}$ group is not
asymptotically free.
For $N \le 5$, it is sufficient to assume that the in the fundamental theory
$\La_{\rm SU} \gg \La_{\rm Sp}$.
In the effective theory, the $\SU{2N - 1}$ gauge group has 2 flavors,
and if the fields $Q'$, $\bar{U}'$, and $\bar{D}'$ were flat directions,
the model would have a runaway \susc vacuum where these fields are
infinite.
The $D$-term potential does not allow these fields to run away, and so
there is no \susc vacuum in this region of moduli space.
(This is the same mechanism that operates in the 3--2 model, but
the present model has classical flat directions.)
Since we have explored all regions of the classical moduli space, we
conclude that \susy is broken in this theory.

We would like to know whether there are local minima on the lepton
branch of the moduli space, and if so, whether these have lower
energy than the local minimum found on the baryon branch.
For $\La_{\rm SU} \gg \La_{\rm Sp}$ 
we can show that the only
minimum is the one found on the baryon branch above.
The reason is simply that if we minimize the energy with $\avg{L}$
held fixed, the energy depends only on the scale $\La_{\rm SU,eff}$
where the $\SU{2N - 1}$ gets strong.
The scale at which the unbroken $\SU{2}$ gauge group becomes
strong is irrelevant, because we have seen that \susy is broken in
the limit where we ignore the non-perturbative effects of the $\SU{2}$
gauge interactions.
The scale $\avg{L}$ appears in the effective theory through the scale
$\La_{\rm SU,eff}$, but otherwise it only controls the size of
higher-dimension operators that give only small corrections to
the vacuum energy.
Therefore, we expect that the vacuum energy as a function of $\avg{L}$ is
$V(\avg{L}) \sim \left| \La_{\rm SU,eff}(\avg{L}) \right|^4$.
This grows with $\avg{L}$, and so we do not expect a vacuum for large
$\avg{L}$.
The analysis above breaks down for $\avg{L} \sim \La_{\rm SU}$.
For $\avg{L} \ll \La_{\rm SU}$, we can use the confined description of the
$\SU{2N - 1}$ dynamics of the previous subsection, so the only remaining
possibility is a vacuum with $\avg{L} \sim \La_{\rm SU}$.
However, in this case, we expect the vacuum energy to be of order
$|\La_{\rm SU}|^4$, which is larger than the vacuum energy \Eq{vacf}
found on the baryon branch.
We conclude that the global minimum of this theory is on the
baryon branch.

The case where $\La_{\rm Sp} \gg \La_{\rm SU}$ appears to be more
complicated, and we cannot rule out the possibility that the global
minimum is on the lepton branch in that case.

For $N \le 5$ and $\La_{\rm Sp} \gg \La_{\rm SU}$, we have not explicitly
shown that there is no \susc vacuum on the lepton branch.
However, we have examined the entire moduli space for
$\La_{\rm SU} \gg \La_{\rm Sp}$ and shown that there is no \susc vacuum.
If there were a \susc vacuum in the limit $\La_{Sp} \gg \La_{\rm SU}$,
there would have to be a critical condition on the
interaction scales $\La_{Sp}$ and $\La_{\rm SU}$ that gave the critical
values at which the \susc vacua are lifted.
However, the moduli space of \susc vacua structure is a
holomorphic function of $\La_{Sp}$ and $\La_{\rm SU}$ \cite{holomorphic}
and so the critical conditions must be holomorphic functions of
$\La_{Sp}$ and $\La_{\rm SU}$.
This means there can be no critical lines
in the space of gauge couplings separating a phase where \susy is
broken from a phase where it is unbroken \cite{nophase}.
This means that \susy is broken also in the limit
$\La_{\rm Sp} \gg \La_{\rm SU}$.

\subsection{The \Sp{4} $\times$ \SU{3} Model}
We now consider the special case $N = 2$, where the superpotential \Eq{Wdyn}
is
\beq[W2]
W_{\rm eff} \sim ( \bar{U}^{3} )^{1/3}.
\eeq
The vacuum is forced away from the origin for small $U$, but the potential
becomes constant for $\avg{ \bar{U}} \gg \La_{\rm SU}$.
The location of the true vacuum therefore depends on the form of the
\Kahler potential.
Yukawa couplings give corrections to the \Kahler potential that
push the field to the origin of moduli space, while gauge corrections do
the reverse.
Since $\Sp{4}$ is asymptotically free, the contribution from the
Yukawa coupling will dominate for large
$\bar{U}$, while the \Sp{4} gauge contributions will dominate for small
$\bar{U}$.
For a range of couplings, there is a \susy-breaking vacuum at large field
values where the theory is fully calculable.
This is an instance of the inverted hierarchy mechanism \cite{invert}
similar to the ones in \Refs{hitoshidirect,CERNdirect}.

\section{Composite Messenger Models}
In this Section, we consider realistic models of gauge-mediated \susy
breaking based on the models analyzed in Section 2.
The model we consider is based on the $N = 3$ model of the previous
section.
This has a global $\SU{5}$ symmetry into which we embed the standard
model gauge group $\SU{3}_C \times \SU{2}_W \times \U1_Y$ in the usual
way.
(We refer to this embedding as $\SU{5}_{\rm SM}$ for brevity.)
The gauge group is therefore
\beq
\Sp{6} \times \SU{5} \times \SU{5}_{\rm SM}
\eeq
with matter content
\beq\bal
Q &\sim (\Yfund, \Yfund, \Yone),
\\
L &\sim (\Yfund, \Yone, \Yfund),
\\
\bar{U} &\sim (\Yone, \bar{\Yfund}, \bar{\Yfund}),
\\
\bar{D} &\sim (\Yone, \bar{\Yfund}, \Yone),
\\
D &\sim (\Yone, \Yone, \Yfund),
\eal\eeq
and a tree-level superpotential
\beq
W = \la Q L \bar{U} + \frac{1}{M^3} (\bar{U}^4 \bar{D}) D.
\eeq
This model differs from the models analyzed above only in that it
contains an additional field $D$ (which cancels the standard-model
anomalies) and there is a higher-dimension term in the tree-level
superpotential.
These new features
are important for the phenomenology of the model, but
they do not affect the qualitative features of the $\Sp{6} \times \SU{5}$
gauge dynamics discussed above.
This model therefore has a \susy-breaking vacuum with
\beq
\avg{\bar{U}} \simeq \frac{\La_{\rm SU}}{4 \pi},
\qquad
F \simeq \avg{F_{\bar{U}}} \simeq
\frac{\la^{5/4} \La_{\rm SU}^{1/4} \La_{\rm Sp}^{7/4}}{4 \pi}.
\eeq
The gauge symmetry is broken in the pattern
\beq
\SU{5} \times \SU{3}_C \times \SU{2}_W \times \U1_Y
\to \SU{3}_C \times \SU{2}_W \times \U1_Y.
\eeq
\Susy breaking is communicated to the standard-model fields via the
messenger pairs $(Q, L)$, $(D, \bar{D})$, and the heavy $\SU{5}$
gauge bosons.
(We are using a Higgs description of the dynamics.)
The masses of the messengers $Q$ and $L$ can be written
\beq\bal
\de\scr{L} &= \myint d^2\th\, M_{QL} Q L + \hc
\\
&\qquad +\, B_{QL} \phi_Q \phi_L + \hc
\\
&\qquad +\, m_Q^2 \phi_Q^\dagger \phi_Q + m_L^2 \phi_L^\dagger \phi_L.
\eal\eeq
Here, $M_{QL}$ is a \susc mass term, $B_{QL}$ is the ``$B$-type'' \susy
breaking mass familiar from traditional gauge-mediated models, and
$m_Q^2$ and $m_L^2$ are soft (non-holomorphic) masses for the messengers.
All of these terms are induced by \susy breaking, and we must estimate
their size.
The \susc and $B$ masses are
\beq
M_{QL} &\simeq \la \avg{\bar{U}}
\simeq \frac{\la \La_{\rm SU}}{4 \pi},
\\
B_{QL} &\simeq \la F.
\eeq
Using \naive dimensional analysis, the soft masses can be estimated
from the gauge exchange diagrams of Fig.~1 to be
\beq
m_Q^2 \simeq m_L^2 \simeq
\frac{g_{\rm SU}^4}{16 \pi^2}
\, \frac{F^2}{\La_{\rm SU}^2} + \cdots
\simeq \frac{16 \pi^2 F^2}{\La_{\rm SU}^2}.
\eeq
The messenger scale that sets the scale for the contributions of these
messengers to the standard model superpartner masses is
\beq
M_{\rm mess} = \frac{B_{QL}}{M_{QL}} \simeq \frac{4\pi F}{\La_{\rm SU}}.
\eeq
We see that $M_{\rm mess}^2 \simeq m_Q^2 \simeq m_L^2$, so the soft mass
contributions to the standard-model superpartner masses are
comparable to the usual gauge-mediated contributions.
The soft mass contribution to the standard-model masses
is not log enhanced from renormalization
group running, since the \susc mass is close to the scale $\La_{\rm SU}$
where the contribution is generated (as long as $\la \sim 1$).

\epsfig{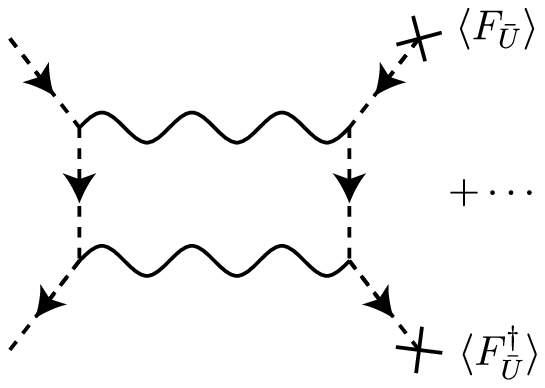}%
{Contributions to the soft scalar mass in the Higgs description.}

In the Higgs picture we are using, the gauge group
$\SU{5} \times \SU{5}_{\rm SM}$ is spontaneously broken down to the
diagonal $\SU{5}$, which we interpret as the low-energy $\SU{5}_{\rm SM}$.
This theory therefore contains gauge messengers, but their contribution
is not calculable because the $\SU{5}$ gauge group is strongly coupled.
(At one loop, the gauge messenger contribution to the scalar masses is
negative \cite{RG}, but there is no reason to believe that this sign
of this result is correct for the strongly coupled case.)
The size of the \susy-breaking masses is the same order as the $Q$ and $L$
messengers discussed above.

The fields $D$ and $\bar{D}$ also act as messengers, and they have
mass terms analogous to those discussed above for $Q$ and $L$.
The \susc and $B$ masses are
\beq[mdd]
M_{D\bar{D}} &\simeq \frac{\avg{\bar{U}}^4}{M^3}
\simeq \frac{1}{M^3} \, \left( \frac{\La_{\rm SU}}{4 \pi} \right)^4,
\\
B_{D\bar{D}} &\simeq \frac{\avg{\bar{U}}^3 F}{M^3}
\simeq \frac{F}{M^3} \, \left( \frac{\La_{\rm SU}}{4 \pi} \right)^3.
\eeq
The $\bar{D}$ soft masses can again be estimated from the
diagrams of Fig.~1 to be
\beq
|m_{\bar{D}}^2|
\simeq \frac{16 \pi^2 F^2}{\La_{\rm SU}^2}.
\eeq
Because $D$ does not feel the strong $\SU{5}$ gauge interactions,
$m_{D}^2 \ll m_{\bar{D}}^2$.
We therefore have
$(B_{D\bar{D}} / M_{D\bar{D}})^2 \simeq m_{\bar{D}}^2$,
so the $\bar{D}$ soft masses are important for communicating
\susy breaking.
In fact, the soft mass contribution is enhanced by renormalization
group evolution from the scale $\La_{\rm SU}$ to the scale
$M_{D\bar{D}}$ where the messengers are integrated out.
This gives a contribution to the squark masses \cite{poppitztrevedi}
\beq[badsquark]
\de m_{\twi{q}}^2 \sim -\left( \frac{g_3^2}{16 \pi^2} \right)^2
m_{\bar{D}}^2 \ln \frac{\La_{\rm SU}}{M_{D\bar{D}}}.
\eeq
This contribution is negative if $m_{\bar{D}}^2 > 0$.
The logarithm cannot be small:
even if $\La_{\rm SU} = M$, the logarithm is of order 10.
It therefore seems sensible to assume that this term dominates.
We see that this model only works if we make the dynamical assumption
$m_{\bar{D}}^2 < 0$.
In this case, we require that $M_{D\bar{D}}^2 \gsim | m_{\bar{D}}^2|$,
so that the \susc mass be large enough that $\avg{\bar{D}} = 0$.
This gives the constraint
\beq
\La_{\rm SU} \gsim 4\pi (M_{\rm mess} M^3)^{1/4}.
\eeq
If we take $M_{\rm mess} \simeq 10$ TeV, and
identify $M$ with the reduced Planck mass
$M_* \simeq 2 \times 10^{18}\GeV$, we have
$\La_{\rm SU} \gsim 7 \times 10^{15}\GeV$.
In order to solve the flavor problem, we want
the the supergravity-mediated contribution to the sparticle mass-squared
$m_{3/2} \sim F / M_*$ to be $\lsim 1\%$ of the gauge-mediated contribution.
This is satisfied for
\beq
\La_{\rm SU} \lsim 2 \times 10^{16}\GeV.
\eeq
We see that if $m_{\bar{D}}^2 < 0$ there is a window where
these models can be realistic even if the scale of the higher dimension
operator is the Planck scale.
For this choice of parameters,
\beq
\sqrt{F} \sim 3 \times 10^{9} \GeV.
\eeq
In this model, the next-lightest supersymmetric particle (NLSP)
will be very long-lived, and may decay late enough in the history of the
universe that its hadronic final states
can induce additional contributions to nucleosynthesis, spoiling the
agreement between the standard theory and experiment.
The authors of \Ref{CERNdirect} obtained a bound of
$\sqrt{F} \lsim 10^{8}\GeV$ from these considerations.
However, this bound is rather model-dependent:
it assumes $R$-parity conservation, and is invalid in inflationary
models with a reheat temperature below the NLSP mass.

Alternatively, if we identify the scale $M$ of the higher-dimension operator
with the grand unification scale, we obtain
$\La_{\rm SU} \gsim 10^{14}\GeV$ and
$\sqrt{F} \gsim (3 \times 10^8 \GeV)$, which may be safe
given the uncertainties involved in these estimates.
In any case, the models will work for sufficiently small $M$.

In these models, the dominant contribution to the standard-model
scalar masses is given by the log-enhanced contribution of the
($D, \bar{D}$) messengers
\beq
m^2_{\tilde{q}} \sim \left( \frac{g^2}{16\pi^2} \right)^2
\left( \frac{4\pi F}{\La_{\rm SU}} \right)^2
\ln \frac{\La_{\rm SU}}{M_{D\bar{D}}},
\eeq
where $M_{D\bar{D}}$ is given by \Eq{mdd}.
The gaugino masses are given by
\beq
m_\la \sim \frac{g^2}{16\pi^2} \frac{4\pi F}{\La_{\rm SU}}.
\eeq
Therefore, in these models the scalar masses are heavier than the
corresponding gaugino masses compared to minimal gauge-mediated models.
However, the minimal gauge-mediation relations between squark and
slepton masses (say) are still satisfied.

\section{Conclusions}
We have discussed a new class of \susy-breaking models based on
direct product groups with a tree-level superpotential.
These models have a large space of flat directions at tree level, but
nonetheless break \susy via the mechanism of s-confinement.
These models have a number of attractive features:
they contain no dimensionful parameters,
and large global symmetries are possible.
By embedding the standard model gauge group in the global symmetry of
a particular model, we have found that a realistic superpartner spectrum
is possible provided that a soft mass term generated by the strong dynamics
is negative.
An interesting direction to explore is to consider a variation of this
model in which the light composite fermions are identified with
standard-model fermions \cite{wip}.

\section{Acknowledgments}
We are grateful to N. Arkani--Hamed for helpful discussions.
M.A.L. thanks R. Rattazzi for helpful discussions, and also
the theory groups at CERN and at LBNL for hospitality during
the course of this work.
J.T. thanks the Aspen Center for Physics, where part of this work
was completed.
M.A.L. was supported by a Sloan Foundation fellowship.
J.T. was supported by the National Science Foundation under grant
PHY-95-14797 and also also partially supported by
the Department of Energy under contract DE-AC03-76SF00098.

\appendix{Appendix: More Supersymmetry Breaking by S-confinement}
In this Appendix, we analyze some additional models related to those
in the main text.
The models have gauge and flavor symmetry group
\beq
G = \Sp{2N} \times SU(5) \times
[ \SU{2N - 1} \times \U1 \times \U1_R ],
\eeq
where the global symmetries are written in brackets.
The matter content is
\beq\bal
Q &\sim (\Yfund, \Yfund) \times (\Yone; 1, 1),
\\
L &\sim (\Yfund, \Yone)
\times (\Yfund; -\sfrac{5}{2N - 1}, -\sfrac{3}{2N - 1}),
\\
\bar{U} &\sim (\Yone, \bar{\Yfund}) \times
(\bar{\Yfund}; -\sfrac{2N - 6}{2N - 1}, \sfrac{2N + 2}{2N - 1}),
\\
\bar{D} &\sim (\Yone, \bar{\Yfund}) \times (\Yone; -6, -12),
\eal\eeq
and there is a tree-level superpotential
\beq
W = \la Q L \bar{U}.
\eeq
For $N = 3$, this is one of the models discussed in the main body of
the text.

The models have been constructed so that the $\Sp{2N}$ factor
has s-confining dynamics.
This can be used to analyze the model for $\La_{\rm Sp} \gg \La_5$,
in a region of moduli space where all \vevs are small compared
to $\La_{\rm Sp}$.
In this regime, the theory has a confined description in terms of
composite chiral superfields.
The effective symmetry group is
\beq
G_{\rm eff} = \SU{5} \times [\SU{2N - 1} \times \U1 \times \U1_R],
\eeq
with matter content (after integrating out massive fields)
\beq\bal
M_{QQ} &= Q Q \sim \Yasymm \times (\Yone; 2, 2),
\\
M_{LL} &= L L \sim \Yone
\times (\kern 1pt\Yasymm\,; -\sfrac{10}{2N - 1}, -\sfrac{6}{2N - 1}),
\\
\bar{D} &\sim \bar{\Yfund} \times (1; -6, -12),
\eal\eeq
with vanishing effective superpotential.
(The $\bar{U}$ equation of motion sets the dynamically generated
superpotential to zero.)
The low-energy theory consists of some singlets, together with a
$\SU{5}$ gauge theory that is believed to break \susy through
non-calculable strong dynamics \cite{fivebarten}.
However, we cannot conclude from this that \susy is broken.
The point is that \susy breaking will induce a non-calculable
potential for the classical flat directions $M_{LL}$, and this potential
may make $M_{LL}$ run away from the origin to a regime where the confined
description is no longer valid.
(In fact, we will show that for $N \ge 3$ the theory has a runaway
\susc vacuum.)
We must analyze the full moduli space of the theory before
we can conclude that \susy is broken.
The analysis differs for various values of $N$, and we proceed on a
case-by-case basis.

\subsection{$N = 1$: Minimal Deconfinement}

This theory has no classical flat directions when the superpotential is
taken into account.
In fact this is the minimal ``deconfined'' description of the
model with gauge group $\SU{5}$ and matter content
$\Yasymm \oplus \bar{\Yfund}$ \cite{fivebarten}.

It is interesting that this theory has a calculable limit.
If we turn off the  $\SU{2}$ gauge coupling, the theory has a classical
moduli space that can be parameterized by the $\SU{2}$ doublets
$M_{Q\bar{D}} = Q \bar{D}$ and $L$, subject to the constraint
\beq
\ep_{\al\be} (L)^\al M_{Q\bar{D}}^\be = 0.
\eeq
Far out along these flat directions, $\SU{2}$ is completely broken,
$\SU{5}$ is broken down to $\SU{4}$, and all fields charged under
$\SU{4}$ are massive.
Gaugino condensation in the $\SU{4}$ Yang--Mills theory generates
a dynamical superpotential for the flat directions
\beq
W_{\rm dyn} \simeq \frac{1}{16\pi^2} (\La_5)^{13/4}
\left( \frac{M_{LL}}{M_{Q\bar{D}}^2} \right)^{1/8}.
\eeq
This superpotential forces $M_{Q\bar{D}}$ to run away to infinity.

If we now turn on an $\SU{2}$ gauge coupling,
all flat directions are lifted at the classical level.
The potential due to $\SU{2}$ gauge couplings is small near the
origin and grows for large fields.
Therefore, for small values of the $\SU{2}$ gauge coupling,
the minimum of the potential will be at large values of
$\avg{M_{Q\bar{D}}}$ and $\avg{L}$, and \susy is broken.
This mechanism for \susy breaking is the same as in the
``3--2 model'' \cite{threetwo}.
We will not analyze this model further.

This analysis proves that there is no \susc vacuum in the parameter region
$\La_5 \gg \La_2$.
However, as discussed in the main text, there can be no phase transitions
as a function of $\La_5 / \La_2$, and so \susy is broken also in the limit
$\La_2 \gg \La_5$, \ie in the original $\SU{5}$ model.

The \susc $\SU{5}$ model has also been related to a calculable model in
Ref.~\cite{murayamasu5} by adding additional vector-like matter and tree-level
superpotential terms, and our conclusions are in agreement.

\subsection{$N = 2$: Supersymmetry Breaking via Supersymmetry Breaking}

The classical flat directions can be parameterized by the
gauge-invariant operator
\beq
M_{LL} = L L \sim (\bar{\Yfund}; -\sfrac{10}{3}, -2).
\eeq

Now consider the effective theory far out along this flat direction.
Na\"\i vely, it appears that the $M_{LL}$ flat direction cannot
be lifted, since the symmetries do not allow a dynamical
superpotential for this field.
However, a careful analysis of the effective theory in this region
of moduli space shows that this argument is not correct because
\susy is broken!

To understand this, note that in terms of the elementary fields,
we are considering vacua with
\beq
\avg{L} = \pmatrix{v & & \cr & v & \cr & & 0 \cr & & 0 \cr},
\eeq
and all other \vevs vanishing.
This breaks $\Sp{4} \to \SU{2}$, and gives a tree-level mass
$\la v$ to two components of $Q$ and $\bar{U}$.
Working out the effective $\SU{2} \times \SU{5}$ gauge theory, one
finds that it has  precisely the matter content of the
theory considered in the previous subsection
(with three additional singlets).
%
As shown above, this theory breaks \susy dynamically,
and this \susy breaking is communicated to the flat fields
$M_{LL}$ by higher-dimension
terms in the effective \Kahler potential.

This model serves as a reminder that an analysis of the flat directions
using the standard arguments based on holomorphy, symmetry, and
classical limits is correct only if the strong sector
of the theory does not itself break \susy.
This subtlety is not present in models with no tree-level
superpotential, since in those theories the effective theory at
a generic point in moduli space is either trivial (the gauge group
is completely broken) or is a pure Yang--Mills theory;
in either case, the low-energy theory does not break \susy.
However, in models with a tree-level superpotential, the classical
equations of motion can force the theory to a singular vacuum
where the unbroken gauge group has charged matter fields.
As illustrated here, such a low-energy theory can break \susy and
invalidate a \naive application of Seiberg's arguments.

\subsection{$N \ge 3$: Supersymmetry Restoring ``Phase Transition"}
The theory has a classical flat directions that can be parameterized by
the gauge-invariants
\beq\bal
M_{LL} = L L
\sim (\Yasymm; -2, -\sfrac{6}{2N - 1}),
\\
\bar{B}_{U} = \bar{U}^4 \bar{D}
\sim (\Yfund; -\sfrac{20N-30}{2N - 1}, \sfrac{-16 N+20}{2N - 1}),
\\
\bar{B}_{D} = \bar{U}^5
\sim (\Yone; -\sfrac{10N-30}{2N - 1}, \sfrac{10 N+10}{2N - 1}).
\eal\eeq
We consider the classical vacuum
\beq
\avg{\bar{U}} = \pmatrix{ u \Yone_5 & 0 \cr},
\qquad
\avg{L} = \pmatrix{ v_1 \Yone_2 & & & & \cr
& \ddots & & & & \cr
& & v_{N - 3} \Yone_2 & & & \cr
& & & \hbox{\bf 0}_4 &\cr
& & & & 0 \cr & & & & 0 \cr}.
\eeq
(The fact that $\avg{L}$ has rank $2(N - 3)$ is enforced by
$\partial W / \partial Q = 0$.)
This breaks $SU(5)$ completely and breaks $\Sp{2N} \to \Sp{6} \times \U1$.
All light matter fields are uncharged under $\Sp{6}$, and
$\Sp{6}$ gaugino condensation gives rise to a dynamical
superpotential
\beq[WPT]
W_{\rm eff} \sim
\left( \frac{\bar{B}_{D}}{M_{LL}^{N - 3}}
\right)^{1 / 4}.
\eeq
For $N \ge 4$, this forces $M_{LL}$ to run away, and there is
a \susc vacuum at infinity.
It may be that there is a local \susy-breaking minimum near the origin,
but we cannot determine this from the present analysis.
Another possibility is that the composite singlet in the s-confined
description has a potential that slopes away from the origin,
and the true vacuum is outside the range of validity of the s-confined
description.

For $N = 3$, both the $\Sp{6}$ and $\SU{5}$ groups confine.
The superpotential \Eq{WPT} is the same as the one discussed
in the main text for the regime $\La_5 \gg \La_6$.
For $\La_6 \gg \La_5$, the analysis in the first part of this Appendix
shows that there is no \susc vacuum for small values of the $\SU{5}$ singlet
fields $M_{LL}$, so we understand how \susy is broken in this case as well.

\end{document}